\documentclass[11pt]{article}

\usepackage[margin=1in]{geometry}
\usepackage{graphicx}
\usepackage{amsmath}
\usepackage{amssymb}
\usepackage{xcolor}
\usepackage{natbib}
\usepackage{url}

\bibpunct{(}{)}{;}{a}{}{,}
\makeatletter
\renewcommand\@biblabel[1]{}
\renewcommand\newblock{\hskip .11em \@plus .33em \@minus .07em\relax}
\newcommand\bibls{\kern.065em\relax}

\makeatother

\definecolor{re180}{HTML}{0072B2}
\definecolor{re550}{HTML}{E69F00}
\definecolor{re1000}{HTML}{009E73}
\definecolor{re2000}{HTML}{56B4E9}
\definecolor{re5200}{HTML}{CC79A7}
\definecolor{re8162}{HTML}{D55E00}

\title{Modeling the wall-normal velocity spectra in channel flow}

\author{Jonathan Massey\textsuperscript{1} and Alexander J. Smits\textsuperscript{2}\thanks{Email address for correspondence: asmits@princeton.edu}\\[2ex]
\small \textsuperscript{1}Center for Turbulence Research, Stanford University, Stanford, CA 94305, USA\\
\small \textsuperscript{2}Dept.\ of Mechanical and Aerospace Engineering, Princeton University, Princeton, NJ 08544, USA}

\date{}

\begin{document}
\maketitle

\begin{abstract}
A model is proposed for the wall-normal Reynolds stress and its one-dimensional spectrum in channel flows.  The model is tested for friction Reynolds numbers from 550 to 5200.
The formulation for the stress is a simple modification of the one devised for boundary layers by \cite{Kunkel2006}, and holds for $y/\delta < 0.75$.  The model for the spectrum holds for all wall distances. It is constructed in wavenumber space, and comprises two principal contributions to the spectrum: energy-containing motions and viscous-dominated motions.  It is in accord with the attached eddy model where that applies, and the viscous contribution scales on the Kolmogorov length and velocity.  Its extension to boundary layers, with no change to the constants, appears promising, including at Reynolds numbers well above those of the channel data.

\end{abstract}

\section{Introduction}\label{sec:intro}

In considering wall-bounded turbulent flows, it is the streamwise component of the time-averaged Reynolds stress, $\overline{{u^2}}$, that has received the most attention.  The reasons for this preference are multifold, but it has the largest magnitude of all the stress components, so that it is a primary agent for turbulent mixing, and it is the component most easily measured, so that its behavior has been extensively recorded.  In contrast, the wall-normal and transverse components, $\overline{{v^2}}$ and $\overline{{w^2}}$ respectively, are much less well documented, probably because they are more difficult to measure.  Here, we are particularly interested in $\overline{{v^2}}$ for three purposes: it plays a crucial role in turbulent transport, exemplified by forming the dominant production term in the equation for the shear stress, and, unlike $\overline{{u^2}}$ and $\overline{{w^2}}$, it does not follow a logarithmic scaling in the mean-flow overlap region  \citep{Townsend1976}.  It also sets the linear (rapid) source of the pressure fluctuations, $-2(\mathrm{d}U/\mathrm{d}y)\,\partial v/\partial x$ for the kinematic pressure, whose streamwise spectrum at each height is $4(\mathrm{d}U/\mathrm{d}y)^2 k_x^2 E_{vv}$.

A semi-empirical formulation for the $\overline{{v^2}}$ profile in boundary layers was proposed by \cite{Kunkel2006}.  Two principal similarity regions were identified: an inner universal region for $y^+ \le 30$ described by $f_{I}$, and an outer region for $y^+ \ge 150$ described by $f_{O}$.  The inner and outer parts were joined using a cubic blend to match the slopes at both ends.
Thus, for $f_{I,O} = \overline{v^{2}}^{\raisebox{-3pt}{\small +}} = \overline{{v^2}}/u_\tau^2 $,
\begin{subequations}
\label{innerouter}
    \begin{align}
        f_{I}(y^+) & = b_1 {y^+}^4 [ (1+ b_2 {y^+}^4) (1+ b_3 {y^+}^4) (1+ b_4 {y^+}^4) (1+ b_5 {y^+}^4) ]^{-{1/4}} \label{inner} \\
        f_{O}(y^+, \delta^+) & = a_3 - V(y^+) (1- y/\delta) - a_3 (y/\delta)^2 (3-2y/\delta), \label{outer}
    \end{align}
\end{subequations}
where $b_1=6.09 \times 10^{-4}$, $b_2=25$, $b_3=1.62 \times 10^{-5}$, $b_4=4.62 \times 10^{-4}$, $b_5=1.96 \times 10^{-7}$, and $a_3=1.78$.  The second term on the right in \eqref{outer} is a viscous correction term with $V(y^+)= 5.58[1-{y^+}^{-0.9}]{y^+}^{-1/2}$, and the third is a wake deviation term.  The outer formulation ensures that in the overlap region $\overline{v^{2}}^{\raisebox{-3pt}{\small +}}$ approaches a universal constant ($a_3$) at sufficiently high Reynolds numbers in accordance with the attached eddy model (AEM) \citep{Townsend1976, Perry1986}.   The overlap region was defined to be where $150< y^+ < 0.2\delta^+$.
The notation is such that $y^+=yu_\tau/\nu$ and the Reynolds number $\delta^+= \delta u_\tau/\nu \, (=Re_\tau)$, where $\delta$ is the  99\% boundary layer thickness or channel half-width, $u_\tau = (\tau_w / \rho)^{1/2}$ is the friction velocity, $\tau_w$ is the shear stress at the wall, and $\rho$ and $\nu$ are the fluid density and kinematic viscosity respectively.

\begin{figure}
    \centering
    \includegraphics[width=0.47\textwidth]{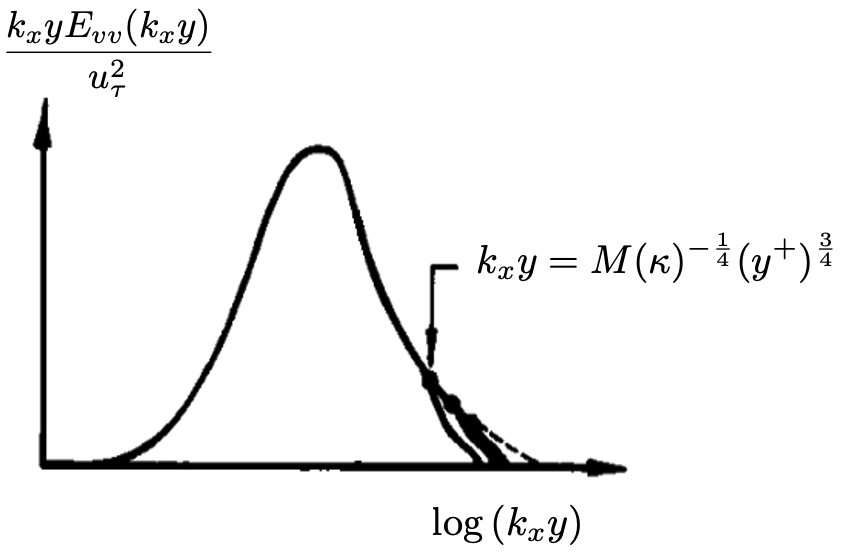} \hspace{4mm}
    \includegraphics[width=0.42\textwidth]{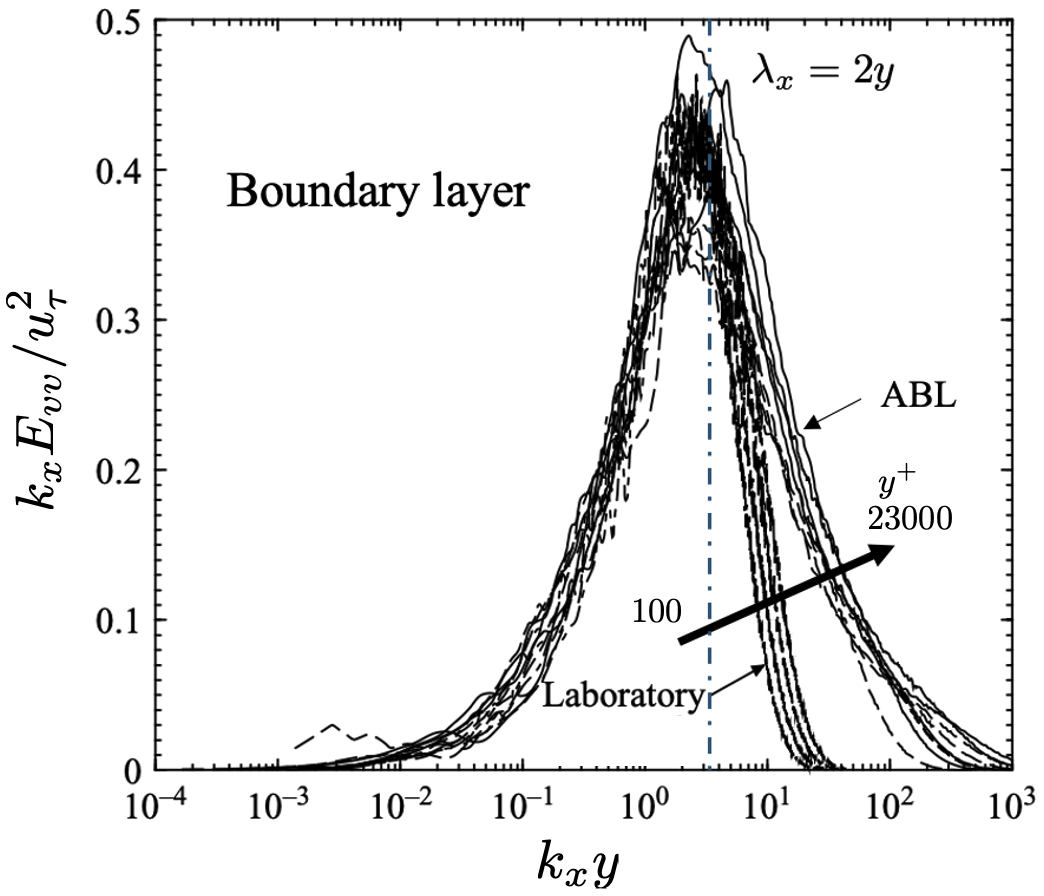}
   \caption{Distributions of $k_xE_{vv}$ in the overlap region.  (a) Attached eddy model, adapted from \cite{Perry1986}.  (b) Laboratory boundary layer results for $\delta^+= 1350$ and atmospheric boundary layer (ABL) for $2.3 \times 10^6$, adapted from \cite{Kunkel2006}. }
    \label{blspectra}
\end{figure}

As to the one-dimensional spectrum $E_{vv}$, according to the AEM it is expected to scale on the wavenumber $k_xy=2\pi y/\lambda_x$ in the overlap region, where  $\lambda_x$ is the streamwise wavelength.
The form expected for the premultiplied spectrum $k_xE_{vv}$ in this region is shown in figure~\ref{blspectra}, together with experimental results for boundary layers at two widely varying Reynolds numbers.  Since $\overline{v^2} = \int_0^\infty k_x E_{vv} \, d(\ln{k_x})$, equal areas in the premultiplied spectrum represent equal energy content.  The measurements are in broad agreement with the AEM, in that the peaks in the spectra correspond to $\lambda_x \approx 2y$, demonstrating the expected linear growth of wavelength with wall distance.  In the atmospheric surface layer, \cite{Kaimal1972} collapsed velocity spectra, including that of the vertical component, as functions of frequency normalized by $U/y$, where $U$ is the local mean velocity, so that in neutral conditions the spectral peak also scales with height.
In addition, the shapes collapse reasonably well at low wavenumbers (large-scale, low-frequency motions), but the fractional energy content at large wavenumbers (small-scale, high-frequency motions) increases with increasing $y^+$ as the length of the inertial region grows (figure~\ref{blspectra}b).  \cite{Perry1986} indicated that the end of the inertial -5/3 region is marked by
\begin{equation}
    k_x y =  M(\kappa)^{-{1 \over 4}}(y^+)^{{3 \over 4}}
\label{Mky}
\end{equation}
(see figure~\ref{blspectra}a), where $\kappa$ is the von K\'arm\'an constant and $M \approx 0.085$. Moreover, the contribution to the energy by the dissipation part of the spectrum decreases at a rate $C(y^+)^{-{1 \over 2}}$, where $C$ is expected to be a universal constant.

We now examine what modifications  to these boundary layer formulations will be needed for channel flows.  Recently, \cite{heisel2026} showed that $\overline{{v^2}}$ in channels, pipes and boundary layers follows the local total shear stress rather than $u_\tau^2$ alone, and examined its spanwise-wavenumber spectra.  The streamwise-wavenumber spectrum considered here was set aside.
In addition, the AEM gives guidance for the behavior of the spectrum in the overlap region, and \cite{PM95} added wake-type eddies to the attached eddies to compute the Reynolds stresses and their spectra in boundary layers.  At present, however, there is no spectral model for $v$ that covers $y^+<150$ and joins it to the overlap and outer regions.  Here, we construct such a model using a  tool that has proven useful in the study of the streamwise-velocity fluctuations \citep{Smits2010a,Gustenyov2025model}, the wall-pressure fluctuations \citep{massey2026two}, and the wall-shear stress fluctuations \citep{Jaroslawski2026TSFP}.  That is, we represent the pre-multiplied spectrum of the wall-normal velocity fluctuations as the linear sum of two log-normal distributions in wavenumber.  This approach yields both a model spectrum and a deeper understanding of the underlying mechanisms that underpin the overall scaling behavior.

\section{Analysis}

\subsection{Stress formulation}

\begin{figure}
    \centering
    \includegraphics[width=0.5\textwidth]{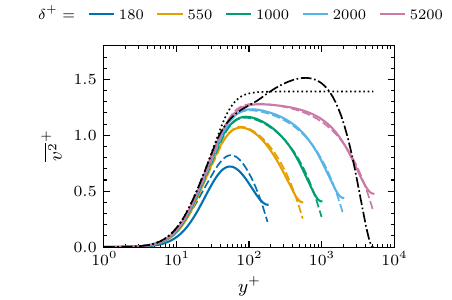}
   \caption{Profiles of $\overline{v^{2}}^{\raisebox{-3pt}{\small +}}$.  Data from DNS by \cite{lee2015direct}: {\color{re180} {\bf --------}}, $\delta^+=180$; {\color{re550} {\bf --------}}, $550$; {\color{re1000} {\bf --------}}, $1000$; {\color{re2000} {\bf --------}}, $2000$; {\color{re5200} {\bf --------}}, $5200$.  Dashed lines, \eqref{channelmodel};
   dotted line, \eqref{inner}; dash-dotted line, model by \cite{Kunkel2006} for $\delta^+=5200$. }
    \label{Kunkel}
\end{figure}

The distributions of $\overline{v^{2}}^{\raisebox{-3pt}{\small +}}$ for channel flow, as obtained by direct numerical simulations \citep{lee2015direct}, are shown in figure~\ref{Kunkel}.  At the lowest Reynolds number (180) there is essentially no overlap region: the smallest attached hierarchy, at $y^+ \approx 100$, already lies beyond $y/\delta = 0.5$ (see also \S\ref{sec:analysis spectra}). At the highest Reynolds number (5200) it is unlikely to have reached an asymptotic state, since the overlap part of the profile is still evolving toward a possible constant plateau value as per the AEM.  In pipe flow, \cite{ZS07} found a region of nearly constant $\overline{v^2}$ for $200 \le y^+ \le 0.1R^+$, where $R$ is the pipe radius, at all but their highest Reynolds numbers.  Nevertheless, the full range is still expected to give a good understanding of Reynolds number trends.

For $\delta^+ =5200$, a comparison is made with the boundary layer model given in \eqref{innerouter}.  Clearly, there is a significant mismatch for $y^+>100$.  However, a simple modification given by
\begin{equation}
    \overline{v^{2}}^{\raisebox{-3pt}{\small +}}(y^+, \delta^+) = b_6 \hat{F}_{I}(y^+) [1-y/(b_6 \delta)]
\label{channelmodel}
\end{equation}
yields excellent agreement with the channel flow data for $\delta^+ \ge 550$ and $y/\delta<0.75$.  Here, $\hat{F}_I$ is the inner function \eqref{inner}, with the boundary layer value $b_2=25$, divided by its plateau value $b_1(b_2b_3b_4b_5)^{-1/4} = 1.39$.  The amplitude $b_6 = 1.10\,(\delta^+)^{0.021}$ is the level to which the overlap part of the profile extrapolates at $y/\delta = 0$, and it is the only coefficient in \eqref{channelmodel} that depends on Reynolds number.  It increases from 1.25 at $\delta^+=550$ to 1.31 at 5200 without limit, and so \eqref{channelmodel} should only be used for $550 \le \delta^+ \le 5200$.  This $b_6$ is the plateau value of \eqref{inner} with $b_2$ replaced by the empirical $64 (\delta^+)^{-0.083}$.  For $y^+>1$, using that $b_2$ in \eqref{inner} itself gives the same profile, since the factor $(1+b_2{y^+}^4)^{-1/4}$ then reduces to $b_2^{-1/4}/y^+$ and only scales the profile.  In \eqref{channelmodel}, $b_6$ takes the place of $a_3$ in \eqref{outer}, and the wake deviation term is modified for channel flow.  Where $\hat{F}_I$ has reached its plateau, \eqref{channelmodel} reduces to $\overline{v^{2}}^{\raisebox{-3pt}{\small +}} = b_6 - y/\delta$.  It falls with $y/\delta$ at the same rate as the total shear stress, $1-y/\delta$, but is offset from it by a constant.  A profile strictly proportional to the shear stress would fall faster by the factor $b_6$, so that the two are told apart by the slope of the profile for $y/\delta > 0.2$.  \cite{heisel2026} scale $\overline{{v^2}}$ on the local shear stress nearer the wall, with a correction that depends on wall distance.  In our model, there does not seem to be a need to include a viscous deviation term.
For $y/\delta>0.75$, the stress approaches a constant value as required by the symmetry of channel flow, where the constant increases slowly with Reynolds number.   To learn more about the underlying scaling, we now examine the spectra.

\subsection{Spectra}\label{sec:analysis spectra}

\begin{figure}
    \centering
    \includegraphics[width=\textwidth]{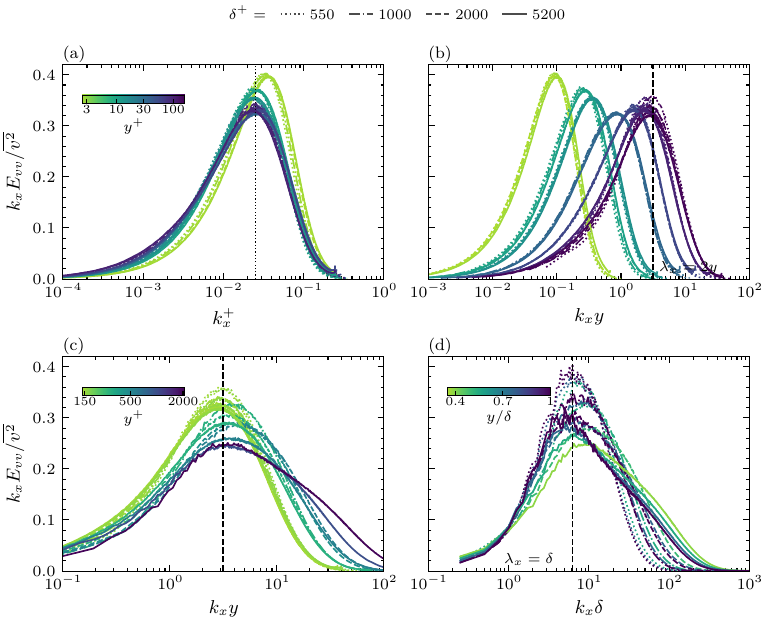}
   \caption{Spectra in channel flow at $\delta^+=550$ to 5200.  (a) Inner region ($2.7<y^+<100$), scaling on $k_x^+$, and (b) scaling on $k_xy$.  (c) Overlap region ($y^+>100$, $y/\delta <0.4$), scaling on $k_xy$.  (d) Outer region ($y/\delta >0.4$), scaling on $k_x \delta$.  Line color gives $y^+$ in (a--c) and $y/\delta$ in (d).}
    \label{nearwall}
\end{figure}

Sample spectra are plotted for the inner, overlap, and outer regions in figure~\ref{nearwall}, using viscous scaling ($k_x^+$), overlap scaling ($k_xy$), and outer scaling ($k_x\delta$), respectively.  The premultiplied spectra are non-dimensionalized by their integral ($=\overline{{v^2}}$) instead of $u_\tau^2$ because it allows a more insightful view of the spectra in the near-wall region, and, as will be shown, it is useful in extending the model for boundary layer flows.
In all three regions, the peaks in the spectra are more or less fixed at a given wavenumber, but all spectra are asymmetric and the distribution of the spectral content changes with $y$ in a way that is specific to each region.

\begin{figure}
    \centering
    \includegraphics[width=\textwidth]{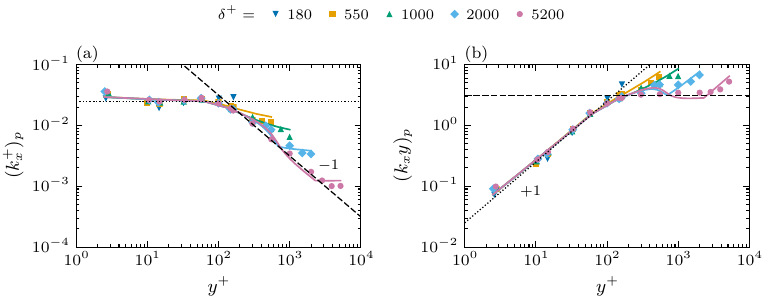}
  \caption{Location of spectral peak.  {\color{re180} $\blacktriangledown$}, $\delta^+=180$; {\color{re550} $\blacksquare$}, $\delta^+=550$; {\color{re1000} $\blacktriangle$}, $\delta^+=1000$; {\color{re2000} $\blacklozenge$}, $\delta^+=2000$; {\color{re5200} {\large $\bullet$}}, $\delta^+=5200$.   (a) Peak in $k_x^+$ scaling.  (b) Peak in $k_xy$ scaling.  Lines ........, $c_1 = 0.025$; - - - - -,  $c_2=\pi$; solid lines in color, peak of the model $(g_1 \! + \! g_2)$ of \S\ref{sec:model} for $\delta^+=550$ to 5200. }
    \label{peak}
\end{figure}

In the inner region ($y^+ \le 100$) the spectra, normalized by $\overline{{v^2}}$, show no Reynolds number dependence, and only a weak dependence on $y^+$ (figure~\ref{nearwall}a).  As indicated in figure~\ref{peak}a, for $y^+>10$ the location of the spectral peak is fixed at $(k_x^+)_p = c_1 \approx 0.025$ (or $(k_xy)_p  =  c_1y^+$, corresponding to $\lambda_x^+ \approx 250$), that is, a length that is approximately twice the height of the first hierarchy in the AEM \citep{Perry1986}.

As $y^+$ increases into the overlap region ($100<y^+ < 0.4 \delta^+$), there is a progression from $k_x^+$ to $k_xy$ scaling (figures~\ref{nearwall}b,c).   At any given wall distance in the overlap region, the distribution of the energy content is almost independent of Reynolds number, but as $y^+$ increases within this region there is a monotonic shift of the energy content to higher wavenumbers at both ends of the spectrum.   The behavior at high wavenumbers is as expected from the AEM (see \eqref{Mky} and figure~\ref{blspectra}b).  At low wavenumbers the energetic part of the spectrum moves with the peak, which migrates from its viscous floor toward $\lambda_x = 2y$ (figure~\ref{peak}b).  \cite{heisel2026} found a dependence on wall distance in the low-wavenumber content of the spanwise-wavenumber spectrum, which they suggest is due to inactive motions. Inactive motions are not expected to contribute significantly to the $v$-spectrum.
The peak location (figure~\ref{peak}b) is now given by $(k_xy)_p  \approx  c_2$  (or $(k_x^+)_p  \approx  c_2/y^+$), where $c_2 \approx \pi$ for $\delta^+=5200$, corresponding to $\lambda_x=2y$.
This observation is in accord with previous work \citep{Kunkel2006} in that  the peak location  in the overlap region is proportional to $y$.  However, only at $\delta^+=5200$ is $(k_xy)_p$ close to constant over an appreciable range ($300<y^+<2850$).  At lower Reynolds numbers the pak location rises continuously with $y^+$, and the larger values (4.7 at $\delta^+=2000$ and 6.7 at $\delta^+=1000$) are only reached for $y/\delta>0.2$, as the outer region is approached.  At $y^+ \approx 300$, $(k_xy)_p = 3.4$ for $\delta^+=2000$, close to $\pi$.  The apparent decrease of $c_2$ with Reynolds number therefore reflects the outer region encroaching on a short overlap region, and $\pi$ may well be close to an asymptotic value, in light of the higher Reynolds number data from turbulent boundary layers shown in figure~\ref{blspectra}b.

As to the outer region ($y/\delta>0.4$), there is a shift from $k_xy$ to $k_x\delta$ scaling (figure~\ref{nearwall}d).   We now see a collapse at the lowest wavenumbers, and a progression with $y/\delta$ at higher wavenumbers.  In pipe flow, \cite{ZS07} likewise found the low-wavenumber part of the spectrum to depend on $y/R$.  At the two highest Reynolds numbers, there is a hint that the peak location $(k_x^+)_p$ returns to a constant value close to $2\pi/\delta^+$ (figure~\ref{peak}a), that is, $\lambda_x \approx \delta$, and the data at the lower Reynolds numbers are in line with the same limit.

The three regions can therefore be combined into a single expression for the peak wavelength,
\begin{equation}
    \lambda_p^+ = \max\left\{ 2\pi/c_1, \; \min\left(2\pi y^+/c_2, \, \delta^+\right) \right\},
\label{peaklaw}
\end{equation}
or equivalently $(k_x^+)_p = \min\{c_1, \, \max(c_2/y^+, \, 2\pi/\delta^+)\}$.  For $c_2 = \pi$ the peak wavelength grows as $2y$ between a viscous floor, $\lambda_x^+ = 2\pi/c_1 \approx 250$, and an outer ceiling, $\lambda_x \approx \delta$.  The transitions follow from the same constants: the peak leaves the floor at $y^+=c_2/c_1 \approx 125$ and reaches the ceiling at $y/\delta=c_2/2\pi=0.5$, close to the region boundaries identified above.  Equation \eqref{peaklaw} describes the data at $\delta^+=5200$.  At $\delta^+=1000$ and 2000 it gives the correct inner and outer limits, but $(k_xy)_p$ rises continuously between them (figure~\ref{peak}b) and approaches the ceiling without reaching it.  For $\delta^+ < 2\pi/c_1 \approx 250$ the ceiling lies below the floor, so the peak remains at the floor and no $k_xy$ range exists, as at $\delta^+=180$.  \cite{pirozzoli2026} bound the streamwise-velocity spectrum between small-scale and large-scale limits in a similar way, and \cite{heisel2026} found a similar progression for the peak of the spanwise-wavenumber spectrum, which follows a dissipation length that is constant near the wall, grows through the overlap region, and changes behavior near $y/\delta=0.5$.

\section{Model Spectrum Formulation}\label{sec:model}

\subsection{Form of the model}

We now construct a model for  $k_xE_{vv}$ in the wavenumber domain. The model has contributions from two eddy populations, broadly representing energy-containing motions ($g_1$) and viscous-scale motions ($g_2$). These distributions overlap in wavenumber space (reflecting the wide range of spectral content in each class of motions), and the energy contained at each wavenumber is taken to be a linear summation of $g_1$ and $g_2$. That is, at any distance from the wall, we propose
\begin{equation}
    k_xE_{vv}/\overline{v^2}= g_1 ( k_x^*, \, y^+, \, \delta^+ ) + g_2(k_x^*, \, y^+, \, \delta^+),
\label{modelsum}
\end{equation}
where $\overline{v^2}$ is used to normalize the spectrum so that its integral is unity, and $k_x^*$ is one of $k_x^+$, $k_xy$, or $k_x\delta$.  We use $k_x^+$ in what follows.  For $g_1$ and $g_2$ we  assume that they can be represented using log-normal distributions in wavenumber, as in
\begin{equation}
    g_{n}=A_{n}\exp \left[ -\left(\frac{\log{k_x^+}-\log{\mu_n^+}}{\log{\sigma_n}} \right)^2\right].
\label{f_general}
\end{equation}
For each $g_n$, therefore, the energy content is distributed around a characteristic wavenumber  $\mu_n^+$, with a width defined by $\sigma_n$ and an amplitude given by $A_{n}$, parameters that are expected to depend on $y^+$ and $\delta^+$.  Here and below, $\log$ denotes the logarithm to base 10.
Each spectrum was fitted by hand to determine these parameters.  The split between the two distributions is less well determined than their sum.

Two populations suffice for $v$, whereas the model spectrum for the streamwise component needs four \citep{Gustenyov2025model}.  The reason is the blocking effect of the wall.  An attached eddy of height $h$ induces streamwise and spanwise motions at a height $y \ll h$, but its wall-normal motion there is reduced in proportion to $y/h$ \citep{Townsend1976}, an effect analysed for shear-free turbulence near a plane surface by \cite{HuntGraham1978}.  Near the wall the large attached eddies therefore contribute mainly the inactive motion of \cite{Bradshaw1967}, which carries little $v$, and the spectrum of $v$ has no $k_x^{-1}$ range, that is, no region of overlap between $y$ and $\delta$ scaling.  Its only overlap is between $y$ and the Kolmogorov scale.  \cite{Perry1986} built their model spectrum of $v$ on this argument, with a spectrum that is constant up to $k_x \sim 1/y$ and follows Kolmogorov scaling beyond.  Here, one population, $g_1$, represents the energy-containing eddies of size comparable with $y$, and the other, $g_2$, the viscous-scale motions, into which the isotropic part of the inertial range is folded.  Blocking removes the $k_x^{-1}$ range but not all of the energy at long wavelengths.  That residual is represented by the low-wavenumber side of $g_1$, which is wider than its high-wavenumber side (see \eqref{sigma1low} below).

\subsection{Widths and peak locations}

First, the widths of $g_1$ and $g_2$ ($\sigma_1$ and $\sigma_2$ respectively) were found to be approximately constant at values of 6 and 3 for all the data considered here.

\begin{figure}
    \centering
    \includegraphics[width=\textwidth]{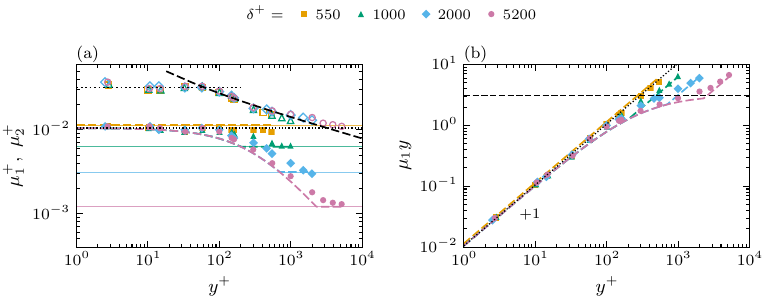}
  \caption{Peak locations of $g_1$ (filled symbols) and $g_2$ (open symbols).  (a) In $k_x^+$ scaling.  ........, $\mu_1^+=0.0105$ and $\mu_2^+=0.032$; {\bf - - - - - -}, decay branch of \eqref{M}; thin horizontal lines in color, $2\pi/\delta^+$.  For \eqref{step}:  {\color{re550} {\bf -- -- -- --}}, $\delta^+=550$; {\color{re1000} {\bf -- -- -- --}}, $\delta^+=1000$; {\color{re2000} {\bf -- -- -- --}}, $\delta^+=2000$; {\color{re5200} {\bf -- -- -- --}}, $\delta^+=5200$.  (b) $\mu_1y$ in $k_xy$ scaling.  ........, $\mu_1^+=0.0105$; {\bf - - - - - -}, $\mu_1y=\pi$, the limit of \eqref{step}; colored dashes as in (a).  Symbols as in figure~\ref{peak}, $\delta^+=550$ to 5200. }
    \label{mucoefficients}
\end{figure}

Next, the characteristic wavenumber $\mu_1$ (that is, the location of the peak value of $g_1$) is plotted in figure~\ref{mucoefficients}.  Its behavior is similar to that of $(k_x^+)_p$ in figure~\ref{peak}a, and it is described by the same three elements as \eqref{peaklaw}: a viscous floor, a ramp that grows as $2y$, and the outer ceiling $\lambda_x = \delta$.  The difference is that the floor and the ramp add rather than hand over, so that
\begin{equation}
    \mu_1^+ = 2\pi/\lambda_1^+, \qquad \lambda_1^+ = \min\left\{\lambda_\nu^+ + 2y^+, \; \delta^+\right\},
\label{step}
\end{equation}
with $\lambda_\nu^+ = 2\pi/0.0105 \approx 600$.  The ramp is that of \eqref{peaklaw} with $c_2 = \pi$, so that no constant is added.  Near the wall $\mu_1^+ = 0.0105$, corresponding to $\lambda_x^+ \approx 600$ (figure~\ref{mucoefficients}a), and it then decreases with $y^+$, as did $(k_x^+)_p$.  Because the floor is added to the ramp, $\mu_1y$ approaches $\pi$ only for $y^+ \gg 300$, so that it is still rising at $\delta^+=5200$ (figure~\ref{mucoefficients}b), whereas the full-spectrum peak $(k_xy)_p$ has already levelled off near $\pi$ (figure~\ref{peak}b).  Near the outer edge, $\mu_1^+$ becomes constant at $2\pi/\delta^+$, that is, $\lambda_x \approx \delta$, which is indicative of $k_x\delta$ scaling and is the same ceiling as for the full spectrum in \eqref{peaklaw}.  The floor and the ceiling coincide at $\delta^+ \approx 600$, so that at $\delta^+=550$ the ceiling applies at all wall distances, within ten per cent of the floor.  At the higher Reynolds numbers the hand-fitted $\mu_1$ approaches the ceiling more gradually than \eqref{step}, which places $g_1$ at a wavelength 10 to 45\% too long for $0.1<y/\delta<0.55$.  A smaller slope would follow the channel data more closely in this range, but \eqref{step} is chosen so that $g_1$ recovers the attached-eddy wavelength $2y$ of \eqref{peaklaw} as $\delta^+$ increases.

The behavior of $\mu_2^+$ is shown by the open symbols in figure~\ref{mucoefficients}a.  Close to the wall, it is approximately constant at a value of 0.032 (corresponding to $\lambda_x^+ \approx 200$), and for $y^+ \gtrsim 60$ it follows \eqref{Mky}.  The two are combined as a ceiling and a decay,
\begin{equation}
    \mu_2^+ = \min\left\{0.032, \; 0.062 \, (\kappa y^+)^{-\alpha_2/4}\right\},
\label{M}
\end{equation}
where $0.062 = \alpha_1 M$ with $\alpha_1=0.73$, and $\alpha_2=\max(0,\, 1-10/y^+)$.  The two branches cross at $y^+ \approx 60$.  Therefore, apart from a correction ($\alpha_2$) that acts only on the second branch, above $y^+ \approx 60$, and tends to unity far from the wall, $\mu_2$ follows \eqref{Mky}, marking the extent of the inertial region and the start of the dissipative region.  The second branch has a simple reading.  With $\epsilon = u_\tau^3/\kappa y$ the Kolmogorov length is $\eta_K^+ = (\kappa y^+)^{1/4}$, so that the second branch of \eqref{M} is \eqref{Mky} with the factor $\alpha_1$, and $g_2$ peaks at an approximately fixed $k_x \eta_K \approx 0.06$.

\subsection{Amplitudes and the area of the model}

\begin{figure}
    \centering
    \includegraphics[width=\textwidth]{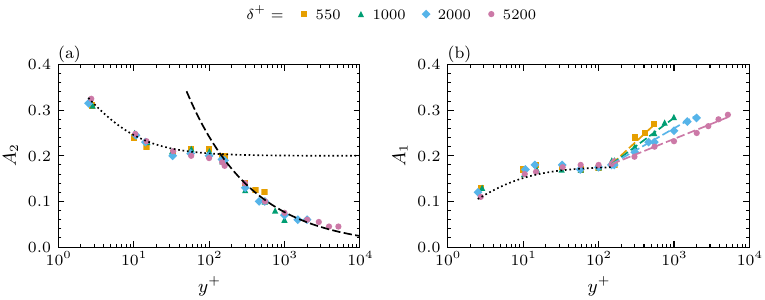}
  \caption{(a) Amplitude of $g_2$: ........, first branch of \eqref{A2}; - - - - -, second branch.  (b) Amplitude of $g_1$: ........, \eqref{A1a}; colored dashes, \eqref{A1b} at each $\delta^+$.   Symbols as in figure~\ref{peak} for $\delta^+=550$ to 5200. }
    \label{amplitude}
\end{figure}

The amplitude of $g_2$, $A_2$, is shown in figure~\ref{amplitude}a.  It can be parameterized in the same way by
\begin{equation}
    A_2 = \min\left\{0.20 + \frac{0.31}{1+y^+/1.7}, \; C(y^+)^{-{1 \over 2}}\right\},
\label{A2}
\end{equation}
where $C=2.42$, and the two branches cross at $y^+ \approx 140$.  The first branch tends to 0.51 at the wall, which would be the value required by \eqref{constraint} below with $I=1$ if $g_1$ were absent, and to 0.20 in the overlap region.  The second branch has the same reading as that of \eqref{M}.  With $\epsilon = u_\tau^3/\kappa y$ the Kolmogorov velocity scale is given by $u_\eta^2/u_\tau^2 = (\kappa y^+)^{-1/2}$.  The share of $\overline{v^2}$ carried by $g_2$ is proportional to $A_2$ (see below), so that, where $\overline{v^{2}}^{\raisebox{-3pt}{\small +}}$ is close to its plateau, the second branch states that the energy of $g_2$ is a fixed multiple of $u_\eta^2$.  With its peak at a fixed $k_x\eta_K$, $g_2$ is then self-similar in Kolmogorov variables.  This is the $(y^+)^{-1/2}$ decrease of the dissipative part of the spectrum noted in \S\ref{sec:intro}.  It does not require an extended $-5/3$ range, which \eqref{Mky} limits to at most about a decade in the overlap region at $\delta^+=5200$.

As to $A_1$, there is a constraint on its magnitude because the integral of \eqref{f_general} yields
\begin{equation}
    (A_1 \log{\sigma_1} +A_2 \log{\sigma_2}) \ln{10} \sqrt{\pi}=I,
\label{constraint}
\end{equation}
where $I=1$ when $(g_1 \! + \!g_2)$ precisely represents $k_xE_{vv}/\overline{v^2}$.  With $\sigma_1=6$ and $\sigma_2=3$, the integrals of the hand-fitted distributions lie between 0.95 and 1.02 for $y^+ < 150$, and between 0.89 and 1.05 for $y^+ >150$, where they vary with $y^+$ and $\delta^+$.  We therefore use
\begin{subequations}
\label{A1}
    \begin{align}
        y^+ <150: \quad  A_1 & = 0.92\,\frac{1 - A_2 \log{\sigma_2} \ln{10}\sqrt{\pi}}{\log{\sigma_1} \ln{10}\sqrt{\pi}}  \label{A1a} \\[1mm]
        y^+ >150: \quad  A_1 & = a_{11}(\log{y^+}/\log{a_{12}}-1){\delta^+}^{a_{13}} + a_{14},   \label{A1b}
    \end{align}
\end{subequations}
where $a_{11}=3.44$, $a_{12}=150$, $a_{13}=-0.37$, and $a_{14}=0.182$  (see figure~\ref{amplitude}b).  Below $y^+=150$, $A_1$ is therefore a fixed fraction, 0.92, of the value that would give $I=1$ in \eqref{constraint}.  The two parts of \eqref{A1} differ by 2\% at $y^+=150$.  With \eqref{A1} and a symmetric $g_1$ the model does not integrate to one.  In the overlap region at $\delta^+=5200$ it lacks a tenth of $\overline{v^2}$, most of it below the peak of $g_1$.  We therefore keep the amplitudes of \eqref{A1}, and with them the variation of $A_1$ with Reynolds number for $y^+>150$, and give $g_1$ a separate width $\sigma_1^-$ for $k_x^+<\mu_1^+$.  Above the peak $\sigma_1=6$ as before.  The integral of $g_1$ is then $A_1 \tfrac{1}{2}(\log{\sigma_1^-}+\log{\sigma_1})\ln{10}\sqrt{\pi}$, and $I=1$ in \eqref{constraint} gives
\begin{equation}
    \log{\sigma_1^-} = \frac{2\,(1 - A_2 \log{\sigma_2} \ln{10}\sqrt{\pi})}{A_1 \ln{10}\sqrt{\pi}} - \log{\sigma_1}.
\label{sigma1low}
\end{equation}
No constant is added.  For $y^+<150$, \eqref{A1a} and \eqref{sigma1low} give $\log{\sigma_1^-} = (2/0.92-1)\log{\sigma_1}$, so that $\sigma_1^- = 8.2$ at every wall distance and Reynolds number.  Above, $\sigma_1^-$ is widest, about 10, in the overlap region at $\delta^+=5200$, and it narrows toward the centerline, to between 5 and 7.  With unit area the shares of $\overline{v^2}$ carried by the two populations are $f_2 = A_2 \log{\sigma_2} \ln{10}\sqrt{\pi} \approx 1.95 A_2$ and $f_1 = 1-f_2$, and they depend on $y^+$ alone.  Through the unit-area formulation of $g_1$, figure~\ref{amplitude}a is therefore also the distribution of $\overline{v^2}$ between the two populations.  The viscous-scale motions carry most of the energy below $y^+ \approx 8$.  Above $y^+ \approx 140$ their share falls as $(y^+)^{-1/2}$, to about 15\% at $y^+=1000$.

\begin{figure}
    \centering
    \includegraphics[width=\textwidth]{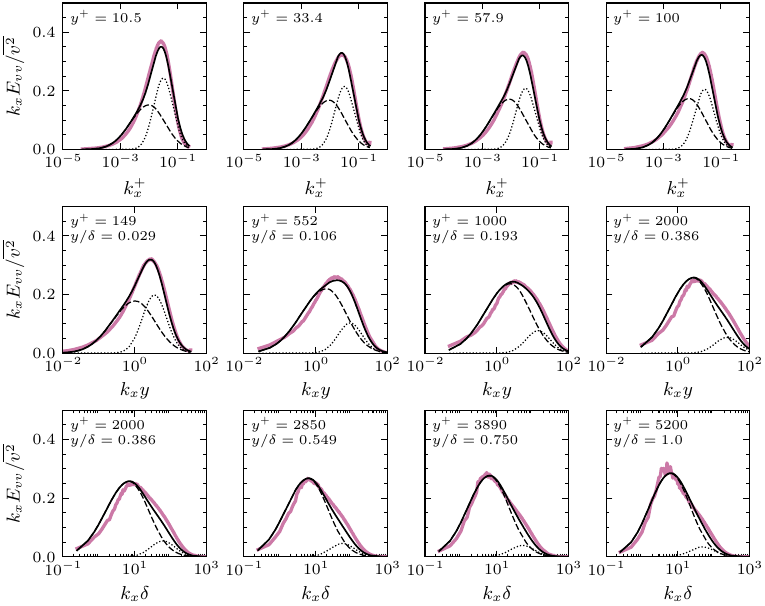}
   \caption{Comparing model prediction to data at $\delta^+=5200$.  Top:  $10.5<y^+<100$, scaling on $k_x^+$.  Middle: $0.029<y/\delta<0.386$, scaling on $k_xy$. Bottom: $0.386<y/\delta<1.0$, scaling on $k_x \delta$.  {\color{re5200} {\bf ----------}}, data; {\bf - - - - - -}, $g_1$; ........, $g_2$; {\bf ----------}, $(g_1 \! + \!g_2)$.}
    \label{modelspectra2}
\end{figure}

\section{Comparison with data}\label{sec:comparison}
\subsection{Channel data}

Our model for the pre-multiplied spectrum of channel flow thus comprises \eqref{modelsum} and \eqref{f_general}, with $\mu_1$ from \eqref{step}, $\mu_2$ from \eqref{M}, $A_2$ from \eqref{A2}, $A_1$ from \eqref{A1}, $\sigma_1=6$ above the peak of $g_1$ and $\sigma_1^-$ from \eqref{sigma1low} below it, and $\sigma_2=3$.  As demonstrated in figure~\ref{modelspectra2} for $\delta^+=5200$, the agreement with data is  generally very good.  Over the stations of figure~\ref{nearwall}, the rms deviation of the model from the normalized spectra is 0.013 to 0.016 at every Reynolds number, about 4\% of the peak value.  It is largest, up to 9\%, for $0.2<y/\delta<0.4$ at $\delta^+=2000$ and 5200, where \eqref{step} places $g_1$ at too long a wavelength, and up to 8\% in the outer region at $\delta^+=550$.

In the inner region the peak of $(g_1 \! + \! g_2)$ is almost independent of $\delta^+$, and for $10<y^+<100$ it stays close to $c_1$ without $c_1$ having been imposed (solid lines in figure~\ref{peak}).  As the inner constants were fitted to the same spectra, this is a consistency check, not a prediction.  Elsewhere the model peak follows \eqref{peaklaw} only approximately, which remains the more accurate summary of the data.

\subsection{Boundary layers}

\begin{figure}
    \centering
    \includegraphics[width=\textwidth]{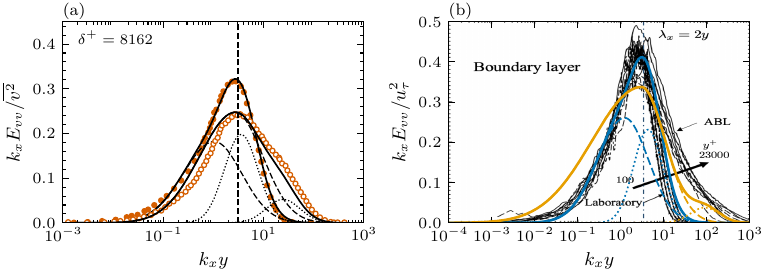}
   \caption{Comparing model prediction to boundary layer experiments.  (a) $\delta^+=8162$, data from \cite{baidya2017distance}: {\color{re8162} {\large $\bullet$}} $y^+=150$; {\color{re8162} {\large $\circ$}} $y^+=2120$ ($y/\delta=0.26$).  (b) Laboratory data at $\delta^+= 1350$ and atmospheric boundary layer (ABL) at $2.3 \times 10^6$, adapted from \cite{Kunkel2006}; the model for {\color{re180} $y^+=196$} and {\color{re550} $y^+=12070$} is scaled by $\overline{v^{2}}^{\raisebox{-3pt}{\small +}}$ from \eqref{outer}.  In (a) each spectrum is normalized by its integral.  {\bf - - - - - -}, $g_1$;  ........, $g_2$;  {\bf ----------}, $(g_1 \! + \! g_2)$.}
    \label{Baidya}
\end{figure}

Preliminary comparisons of our channel flow model with boundary layers appear promising, based on the results  shown in  figure~\ref{Baidya}.   For the \cite{baidya2017distance} data ($\delta^+=8162$) the comparisons with the model are made at $y^+=150$ and 2120.  For the \cite{Kunkel2006} laboratory data ($\delta^+=1350$) the comparison is made at $y^+=196$, and for their ABL data ($\delta^+=2.3 \times 10^6$) at $y^+=12070$; the model is drawn over their figure, scaled by $\overline{v^{2}}^{\raisebox{-3pt}{\small +}}$ from \eqref{outer}.  The \cite{baidya2017distance} spectra are normalized by their own integrals, as in figure~\ref{modelspectra2}.  All the constants derived from channel flow were unchanged for these comparisons, and the widths of $g_1$ follow from \eqref{sigma1low}, as in channel flow.  In particular, \eqref{step} gives $\mu_1y = 3.07$ at the ABL station, close to the attached-eddy value $\pi$, without adjustment.  From figure~\ref{Baidya}, the model reproduces the location and height of the peak to within five per cent at both heights in panel (a), and it lies within the envelope of the \cite{Kunkel2006} laboratory spectra in panel (b).  For the ABL its peak lies about 25\% below the \cite{Kunkel2006} spectra.  At this Reynolds number \eqref{A1b} is used far outside the range in which it was fitted, and \eqref{sigma1low} then gives $\sigma_1^- \approx 40$, which spreads the energy missing from the peak into a long tail at low wavenumbers.  \cite{JH08} found the outer modes of the wall-normal velocity to be stronger in boundary layers than in internal flows, so the outer part of $g_1$ is the part of the model least likely to transfer.

\section{Conclusions}

The wall-normal velocity fluctuations in channel flows were examined in terms of the scaling behavior of the stress $\overline{v^{2}}^{\raisebox{-3pt}{\small +}}$ and its one-dimensional wavenumber spectrum $E_{vv}$.  It was shown that the stress profiles can be accurately described using a simple modification to the formulation originally proposed for boundary layers by \cite{Kunkel2006}, in which only the amplitude $b_6$ depends on Reynolds number.  In addition, we established the scaling behavior of the pre-multiplied spectrum $k_xE_{vv}$ over the inner region, overlap region and outer region, thereby providing an account of the streamwise-wavenumber spectrum over the full width of the shear layer.

The peak wavelength is summarized by \eqref{peaklaw}.  It grows as $2y$, as proposed by the attached eddy model, between a viscous floor fixed in wall units and an outer ceiling at $\lambda_x \approx \delta$, so that no range of $k_xy$ scaling can exist below $\delta^+ \approx 250$, and a plateau in $(k_xy)_p$ is demonstrated only at $\delta^+=5200$.  Through the overlap region the energy shifts to higher wavenumbers at both ends of the spectrum, at high wavenumbers because the inertial region lengthens, and at low wavenumbers because the peak migrates from the floor toward $2y$.

A model for the spectrum consisting of the linear sum of two log-normal distributions in wavenumber, $g_1$ and $g_2$, was found to represent its dependence on wall distance and Reynolds number to within a few per cent of the peak value.  The distribution $g_2$ represents the viscous-scale motions.  Close to the wall its peak location is fixed in inner units ($\lambda_x^+ \approx 200$).  Further from the wall its peak location and its energy scale on the Kolmogorov length and velocity, for $y^+>60$ and $y^+>140$ respectively, as anticipated by \cite{Perry1986}.  The rest of the energy is modeled by $g_1$.  Its peak wavelength is the sum of a viscous floor, $\lambda_x^+ \approx 600$, and the ramp $2y$ of the full spectrum, capped at $\lambda_x \approx \delta$.  The floor is between two and three times the wavelength where the full spectrum peaks near the wall.  Because the floor is added to the ramp, $\mu_1y$ is still rising at $\delta^+=5200$, whereas the full-spectrum peak $(k_xy)_p$ has already levelled off near $\pi$.

Together with \eqref{channelmodel} the model supplies the spectrum of the linear source of the wall pressure, $4(\mathrm{d}U/\mathrm{d}y)^2 k_x^2 E_{vv}$, at each height.  The wall pressure also depends on the two-height cross-spectrum of $v$, which a one-dimensional spectrum cannot supply.  \cite{panton2017correlation} fitted a logarithmic growth with Reynolds number to the linear contribution in channel simulations, and \cite{massey_linear_2026} argue for the asymptotic inner scaling behavior of $\partial v/\partial x$ that it becomes Reynolds-number independent.  Which one holds is left to future work.

The extension of this channel flow model spectrum to boundary layers appears promising, based on some preliminary comparisons with data at Reynolds numbers up to $2.3 \times 10^6$.  All the constants derived from channel flow at Reynolds numbers up to 5200 transferred unchanged.  In the ABL the peak of $g_1$ falls close to the attached-eddy wavelength without adjustment, although the peak of the spectrum there is about 25\% low.  Its performance at other locations in the boundary layer,  and a possible extension to pipe flows,  will need to be tested using additional data.   These tests will be the subject of future work.

\paragraph{Acknowledgements.} We would like to thank the authors of the original datasets for making their data openly available, and Mr.\ Kyle Devlin for his helpful comments on an earlier draft.

\paragraph{Funding.} The support from DARPA under award HR0011-24-9-0465 is gratefully acknowledged.

\paragraph{Declaration of interests.} The authors report no conflict of interest.

\paragraph{Author ORCIDs.} J. Massey, \url{https://orcid.org/0000-0002-2893-955X};\\ A. J. Smits, \url{https://orcid.org/0000-0002-3883-8648}.

\bibliographystyle{jfm}
\bibliography{refs}

\begin{thebibliography}{19}
\expandafter\ifx\csname natexlab\endcsname\relax\def\natexlab#1{#1}\fi
\def\au#1{#1} \def\ed#1{#1} \def\yr#1{#1}\def\at#1{#1}\def\jt#1{\textit{#1}}
  \def\bt#1{#1}\def\bvol#1{\textbf{#1}} \def\vol#1{#1} \def\pg#1{#1}
  \def\publ#1{#1}\def\arxiv#1{#1}\def\org#1{#1}\def\st#1{\textit{#1}}

\bibitem[Baidya {\em et~al.\/}(2017)Baidya, Philip, Hutchins, Monty \&
  Marusic]{baidya2017distance}
{\sc \au{Baidya, R.}, \au{Philip, J.}, \au{Hutchins, N.}, \au{Monty, J.~P.} \&
  \au{Marusic, I.}} \yr{2017}  \at{Distance-from-the-wall scaling of turbulent
  motions in wall-bounded flows}.  \jt{Phys. Fluids}  \bvol{29}~(2),
  \pg{020712}.

\bibitem[Bradshaw(1967)]{Bradshaw1967}
{\sc \au{Bradshaw, P.}} \yr{1967}  \at{{`Inactive'} motion and pressure
  fluctuations in turbulent boundary layers}.  \jt{J. Fluid Mech.}  \bvol{30},
  \pg{241--258}.

\bibitem[Gustenyov {\em et~al.\/}(2025)Gustenyov, Bailey \&
  Smits]{Gustenyov2025model}
{\sc \au{Gustenyov, N.}, \au{Bailey, S. C.~C.} \& \au{Smits, A.~J.}} \yr{2025}
  \at{A model spectrum for turbulent wall-bounded flow}.  \jt{J. Fluid Mech.}
  \bvol{1016},  \pg{A23}.

\bibitem[Heisel {\em et~al.\/}(2026)Heisel, Deshpande \& Katul]{heisel2026}
{\sc \au{Heisel, Michael}, \au{Deshpande, Rahul} \& \au{Katul, Gabriel~G.}}
  \yr{2026}  \at{On the wall-normal velocity variance in canonical wall-bounded
  turbulence}.  \jt{J. Fluid Mech.}  \bvol{1033},  \pg{A30}.

\bibitem[Hunt \& Graham(1978)]{HuntGraham1978}
{\sc \au{Hunt, J. C.~R.} \& \au{Graham, J. M.~R.}} \yr{1978}  \at{Free-stream
  turbulence near plane boundaries}.  \jt{J. Fluid Mech.}  \bvol{84}~(2),
  \pg{209--235}.

\bibitem[Jaroslawski {\em et~al.\/}(2026)Jaroslawski, Massey, Cabrera-Booman,
  McKeon, Devlin \& Smits]{Jaroslawski2026TSFP}
{\sc \au{Jaroslawski, T.}, \au{Massey, J.}, \au{Cabrera-Booman, F.},
  \au{McKeon, B.~J.}, \au{Devlin, K.} \& \au{Smits, A.~J.}} \yr{2026}
  Subconvective wall-shear fluctuations in high-{R}eynolds number boundary
  layers.  \bt{In {\em 14th International Symposium on Turbulence and Shear
  Flow Phenomena (TSFP14) Heidelberg, Germany\/}},  \pg{pp. 1--6}.

\bibitem[Jim\'{e}nez \& Hoyas(2008)]{JH08}
{\sc \au{Jim\'{e}nez, J.} \& \au{Hoyas, S.}} \yr{2008}  \at{Turbulent
  fluctuations above the buffer layer of wall-bounded flows}.  \jt{J. Fluid
  Mech.}  \bvol{611},  \pg{215--236}.

\bibitem[Kaimal {\em et~al.\/}(1972)Kaimal, Wyngaard, Izumi \&
  Cot\'{e}]{Kaimal1972}
{\sc \au{Kaimal, J.~C.}, \au{Wyngaard, J.~C.}, \au{Izumi, Y.} \& \au{Cot\'{e},
  O.~R.}} \yr{1972}  \at{{Spectral characteristics of surface-layer
  turbulence}}.  \jt{Q. J. R. Meteorol. Soc.}  \bvol{98},  \pg{563--589}.

\bibitem[Kunkel \& Marusic(2006)]{Kunkel2006}
{\sc \au{Kunkel, G.~J.} \& \au{Marusic, I.}} \yr{2006}  \at{{Study of the
  near-wall-turbulent region of the high-{R}eynolds-number boundary layer using
  an atmospheric flow}}.  \jt{J. Fluid Mech.}  \bvol{548},  \pg{375--402}.

\bibitem[Lee \& Moser(2015)]{lee2015direct}
{\sc \au{Lee, M.} \& \au{Moser, R.~D.}} \yr{2015}  \at{Direct numerical
  simulation of turbulent channel flow up to ${Re}_\tau \approx 5200$}.  \jt{J.
  Fluid Mech.}  \bvol{774},  \pg{395--415}.

\bibitem[Massey {\em et~al.\/}(2026{\natexlab{{\em a\/}}})Massey, Smits \&
  McKeon]{massey2026two}
{\sc \au{Massey, J. M.~O.}, \au{Smits, A.~J.} \& \au{McKeon, B.~J.}}
  \yr{2026{\natexlab{{\em a\/}}}}  \at{Two-component inner--outer scaling model
  for the wall-pressure spectrum at high {R}eynolds number}.  \jt{J. Fluid
  Mech.}  \bvol{1034},  \pg{A60}.

\bibitem[Massey {\em et~al.\/}(2026{\natexlab{{\em b\/}}})Massey, Zimmerman,
  Klewicki \& McKeon]{massey_linear_2026}
{\sc \au{Massey, J. M.~O.}, \au{Zimmerman, S.~J.}, \au{Klewicki, J.~C.} \&
  \au{McKeon, B.~J.}} \yr{2026{\natexlab{{\em b\/}}}} An inner-scaled linear
  contribution to wall-pressure variance at high {R}eynolds number.
  arXiv:2607.02395.

\bibitem[Panton {\em et~al.\/}(2017)Panton, Lee \&
  Moser]{panton2017correlation}
{\sc \au{Panton, R.~L.}, \au{Lee, M.} \& \au{Moser, R.~D.}} \yr{2017}
  \at{Correlation of pressure fluctuations in turbulent wall layers}.
  \jt{Phys. Rev. Fluids}  \bvol{2}~(9),  \pg{094604}.

\bibitem[Perry {\em et~al.\/}(1986)Perry, Henbest \& Chong]{Perry1986}
{\sc \au{Perry, A.~E.}, \au{Henbest, S.~M.} \& \au{Chong, M.~S.}} \yr{1986}
  \at{A theoretical and experimental study of wall turbulence}.  \jt{J. Fluid
  Mech.}  \bvol{165},  \pg{163--199}.

\bibitem[Perry \& Marusic(1995)]{PM95}
{\sc \au{Perry, A.~E.} \& \au{Marusic, I.}} \yr{1995}  \at{A wall-wake model
  for the turbulence structure of boundary layers. {P}art 1. {E}xtension of the
  attached eddy hypothesis}.  \jt{J. Fluid Mech.}  \bvol{298},  \pg{361--388}.

\bibitem[Pirozzoli \& Hwang(2026)]{pirozzoli2026}
{\sc \au{Pirozzoli, S.} \& \au{Hwang, Y.}} \yr{2026}  \at{A spectral model for
  the wall-normal distribution of velocity variance in turbulent wall layers}.
  \jt{J. Fluid Mech.}  \bvol{1039},  \pg{A27}.

\bibitem[Smits(2010)]{Smits2010a}
{\sc \au{Smits, A.~J.}} \yr{2010} {High Reynolds number wall-bounded turbulence
  and a proposal for a new eddy-based model}.  \bt{In {\em Notes on Numerical
  Fluid Mechanics and Multidisciplinary Design\/} (ed. \ed{M.~Deville,
  Thi\^en-Hi\^ep \& P.~Sagout})}, ,  \vol{vol. 110},  \pg{pp. 51--62}.
  \publ{Springer}.

\bibitem[Townsend(1976)]{Townsend1976}
{\sc \au{Townsend, A.~A.}} \yr{1976} {\em The Structure of Turbulent Shear
  Flow\/}.  \publ{Cambridge, UK: Cambridge University Press}.

\bibitem[Zhao \& Smits(2007)]{ZS07}
{\sc \au{Zhao, R.} \& \au{Smits, A.~J.}} \yr{2007}  \at{Scaling of the
  wall-normal turbulence component in high-{R}eynolds-number pipe flow}.
  \jt{J. Fluid Mech.}  \bvol{576},  \pg{457--473}.

\end{thebibliography}

\end{document}